**Chapter**

# Superconducting Parametric Amplifiers: Resonator Design and Role in Qubit Readout

*Babak Mohammadian*

## Abstract

Superconducting parametric amplifiers (SPAs) are critical components for ultra-low-noise qubit readout in quantum computing, addressing the critical challenge of amplifying weak quantum signals without introducing noise that degrades coherence and computational fidelity. Unlike classical amplifiers, SPAs can achieve or closely approach quantum-limited performance, specifically the *Standard Quantum Limit* (SQL) of half a photon of added noise for phase-preserving amplification. The core principle of SPAs relies on parametric amplification, where energy is transferred from a strong pump tone to a weak input signal through non-dissipative nonlinear mixing processes. This is enabled by intrinsic nonlinearities in superconducting materials, primarily kinetic inductance in thin films (e.g., NbTiN, Al) and, more significantly, the Josephson effect in Josephson junctions. These nonlinear elements facilitate frequency mixing (three-wave or four-wave mixing) and can operate in phase-preserving or phase-sensitive amplification modes, with the latter allowing for noise squeezing below the SQL. This chapter emphasizes the significant role of resonator design in determining critical SPA performance metrics such as gain, bandwidth, and noise characteristics. It details both lumped-element (LC) and distributed-element (coplanar waveguide, CPW) resonators, discussing their unique properties, suitability for different frequency ranges, and the importance of achieving high-quality factors ($Q$) for efficient energy storage and minimal loss. A practical design and simulation of a meandered quarter-wavelength CPW resonator coupled to a feed line is presented, illustrating how precise control over geometric parameters optimizes resonant frequency, coupling strength, and quality factor for high-fidelity qubit state discrimination.

## 1. Introduction

Reading out quantum states with high precision while minimizing added noise remains one of the foremost challenges in quantum computing. Qubits are inherently fragile and can be easily perturbed by unwanted interactions, particularly during measurement. This noise degrades coherence and reduces the computational fidelity.

Therefore, ultra-low-noise amplification is essential to preserve quantum information and improve readout accuracy [1, 2].

Superconductivity provides ultra-low-noise quantum measurements. Defined by zero electrical resistance and the Meissner effect [3], superconductors exhibit extremely low dissipation at cryogenic temperatures. For quantum circuits, operating at millikelvin temperatures is critical—not simply to achieve superconductivity, but to suppress thermal excitations and maintain quantum coherence. These conditions enable high-quality factors (Q) in microwave resonators used for qubit control and readout [4, 5]. Modern superconducting processors employ thin-film fabrication to realize complex circuits containing Josephson junctions and high-Q resonators [6], which provide low-loss energy storage and nonlinear interactions essential for quantum operations [7, 8].

In superconducting processors, qubit signals are extremely weak—on the order of only a few photons at microwave frequencies (4–8 GHz). Detecting these signals without excessive added noise is crucial for high-fidelity state discrimination. Consequently, the first amplification stage must provide high gain, minimal added noise, and ideally, quantum-limited performance [9, 10]. Resonator design within parametric amplifiers strongly influences these metrics, as geometry and material properties dictate bandwidth, gain flatness, and noise.

Traditional amplifiers such as High Electron Mobility Transistors (HEMTs) [11], even at cryogenic temperatures, exhibit noise temperatures in the 1–5 K range [12, 13]. As classical semiconductor devices, HEMTs are constrained by thermal and transistor noise mechanisms and cannot approach quantum-limited performance. For quantum systems, this added noise is prohibitive: Weak microwave signals corresponding to a few photons are easily buried under several kelvins of amplifier noise. By contrast, quantum-limited amplification requires an added noise of only about half a photon at the signal frequency—the Standard Quantum Limit (SQL)—the minimum noise permitted by quantum mechanics for a phase-preserving amplifier [14, 15]. Noise temperature here quantifies the equivalent thermal noise power introduced by the amplifier, providing an intuitive comparison of noise performance.

Superconducting parametric amplifiers (SPAs) can closely approach the quantum limit, making them indispensable for quantum measurement. They function by transferring energy from a strong pump tone to a weak input signal through nonlinear mixing. This process enables amplification with minimal added noise, particularly when using superconducting materials such as NbTiN and Al to utilize nonlinearities arising from kinetic inductance or the Josephson effect [16, 17]. The nonlinear inductance of Josephson junctions is especially critical, as it directly enables parametric amplification. In superconducting qubit platforms, SPAs are integrated into the readout chain, where the qubit state is dispersively encoded in a resonator signal that must be amplified and distinguished from noise [18, 19]. Resonator design in SPAs therefore plays a central role in setting performance metrics, including the achievable signal-to-noise ratio (SNR).

Moreover, parametric amplifiers enable quantum-limited phase-sensitive detection [20] and noise squeezing [21], which can suppress noise in one quadrature below the SQL. This capability is particularly valuable in quantum error correction [22] and quantum sensing applications [23], where maximizing information extraction while minimizing disturbance is essential.

This chapter focuses on the design and operation of superconducting parametric amplifiers, with emphasis on resonator-based architectures. We examine how resonator geometry and material properties determine key performance metrics such as gain, bandwidth, and noise. We also discuss the role of SPAs in dispersive qubit readout chains and their impact on the achievable signal-to-noise ratio (SNR). By combining theoretical foundations with practical design considerations, the chapter provides a comprehensive perspective on how resonator design enables near-quantum-limited amplification in superconducting processors.

## 2. Superconducting parametric amplification

A parametric amplifier consists of three fundamental components. The first is a nonlinear element, which is essential for approaching the Standard Quantum Limit (SQL). To achieve near-quantum-limited performance, the amplifier must employ a non-dissipative nonlinearity that enables frequency mixing without adding excess noise. In contrast to linear systems, where signals propagate without modifying one another, nonlinear systems allow interactions that generate new frequency components and complex dynamics [17]. Such nonlinearities can arise from current-dependent inductances [24], voltage-dependent capacitances, or quantum effects in Josephson junctions [25, 26].

The second component is the pump: a strong microwave or RF tone that supplies the external energy driving the parametric process [10]. Finally, a resonant circuit is typically employed to enhance the interaction between the signal and the nonlinear element, thereby improving gain and frequency selectivity by confining the operation to well-defined modes [27].

### 2.1 Superconductivity and kinetic inductance

Superconductivity arises when a material is cooled below its critical temperature $T_c$, a state characterized by zero electrical resistance and expulsion of magnetic fields —the Meissner effect [28]. Within BCS theory, this phase emerges from Cooper pairs, electron pairs bound *via* phonon-mediated attraction, which condense into a macroscopic quantum state [29]. Below $T_c$, an energy gap $\Delta(T)$ opens at the Fermi surface, suppressing scattering and enabling dissipationless current flow.

The electromagnetic response of superconductors is described by the London equations [30]:

$$\frac{\partial \vec{J}_s}{\partial t} = \frac{1}{\mu_0 \lambda^2} \vec{E}, \qquad \nabla \times \vec{J}_s = -\frac{1}{\lambda^2} \vec{H}, \tag{1}$$

where $\lambda$ is the London penetration depth. These equations capture the essential features of perfect conductivity and magnetic field expulsion.

Despite having zero DC resistance, superconductors exhibit a finite *kinetic inductance* at nonzero frequencies, arising from the inertia of Cooper pairs [31]. When a time-varying current flows, Cooper pairs must be accelerated, requiring energy and producing a reactive impedance that stores energy in carrier motion rather than magnetic fields. For small currents, this inductance remains nearly constant; at higher currents, however, pair-breaking induces nonlinear behavior:

$$L_K(I) = L_K(0)\left(1 + \frac{I^2}{I_*^2} + \cdots\right), \tag{2}$$

where $I_*$ is a characteristic current scale [24, 32]. This intrinsic nonlinearity makes kinetic inductance a valuable mechanism for superconducting parametric amplification.

### 2.2 Josephson inductance

Josephson inductance provides another nonlinear mechanism employed in parametric amplification [33]. It originates from the Josephson junction, which consists of two superconducting electrodes separated by a nanometer-scale insulating barrier. Although separated by this barrier, Cooper pairs can tunnel quantum mechanically from one electrode to the other without resistance [25], as illustrated in **Figure 1**.

The supercurrent through the junction follows the first Josephson relation:

$$I = I_c \sin \phi, \tag{3}$$

where $I_c$ is the critical current and $\phi$ is the superconducting phase difference. This nonlinear current-phase relation underpins the use of Josephson junctions in nonlinear circuits and parametric amplifiers.

The second Josephson relation links the time evolution of the phase to the voltage across the junction:

$$\frac{d\phi}{dt} = \frac{2e}{\hbar} V \quad \text{or equivalently} \quad V = \frac{\Phi_0}{2\pi} \frac{d\phi}{dt}, \tag{4}$$

where $\Phi_0 = h/(2e)$ is the magnetic flux quantum. By combining the two Josephson relations and comparing with the definition of an inductor, $V = L_J \, dI/dt$, one obtains:

$$L_J = \frac{\Phi_0}{2\pi I_c \cos \phi}. \tag{5}$$

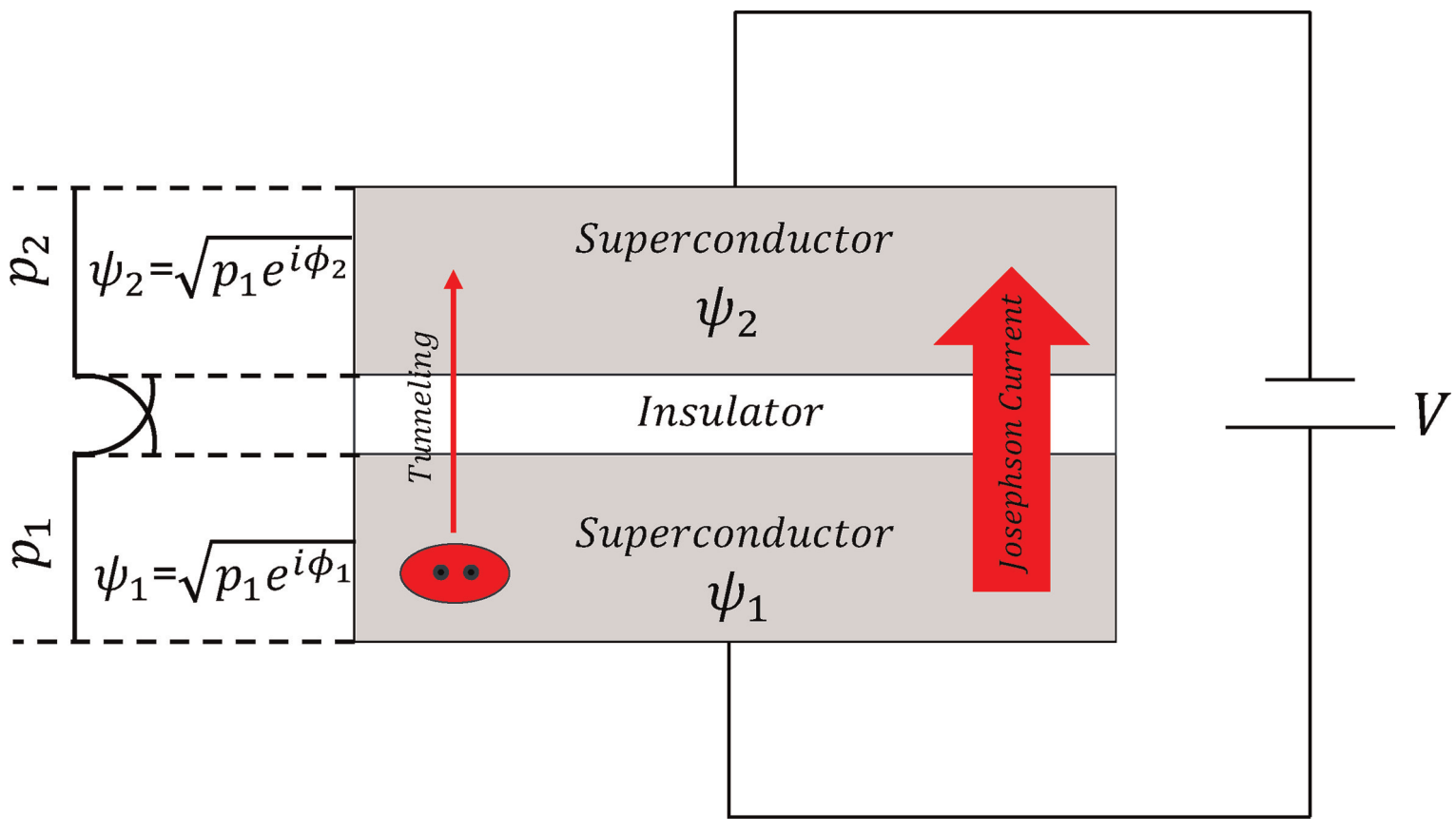

**Figure 1.**
*Josephson junction schematic: two superconductors separated by a thin insulating barrier (a few nanometers), allowing Cooper pairs to tunnel* via *overlapping macroscopic wavefunctions. The superconducting order parameters are represented as* $\psi_1 = \sqrt{p_1}e^{i\phi_1}$ *and* $\psi_2 = \sqrt{p_2}e^{i\phi_2}$*, where the phase difference* $\phi = \phi_2 - \phi_1$ *defines the superconducting phase across the junction.*

This expression shows that the Josephson inductance $L_J$ is both nonlinear and tunable, since it depends explicitly on the phase difference $\phi$ across the junction. As $\phi$ approaches $\pi/2$, $\cos\phi$ vanishes and the inductance diverges, producing strong nonlinearity. Moreover, because $L_J$ is inversely proportional to the critical current $I_c$, its value can be engineered by adjusting junction parameters or controlled dynamically using an external flux bias (**Table 1**).

| Feature | Kinetic inductance | Josephson inductance |
|---|---|---|
| Physical origin | Inertia of Cooper pairs | Quantum tunneling of Cooper pairs |
| Nonlinearity type | Kerr-type (third order) | Depends on circuit design and pumping |
| Key parameter | Current-dependent suppression of gap | Phase difference across junction $\phi$ |
| Mathematical form | $L_k(I) \approx L_{k0}\left(1 + \alpha I^2\right)$ | $L_J = \frac{\Phi_0}{2\pi I_c \cos\phi}$ |
| Tuning mechanism | Fixed by material properties | Tunable *via* magnetic flux in SQUIDs |
| Typical implementation | Superconducting transmission lines | Josephson junctions/SQUIDs |
| Applications | SPAs, kinetic inductance detectors | SPAs, qubits, tunable resonators |

*Note: $\Phi_0$ is the magnetic flux quantum; $I_c$ is the junction critical current. The effective nonlinearity type of Josephson inductance depends on the circuit configuration and pumping scheme. It can be engineered for second-order nonlinearity (e.g., in flux-pumped SQUIDs for 3WM) or third-order (Kerr-type) nonlinearity (e.g., in current-pumped single junctions for 4WM).*

**Table 1.**
*Comparison of kinetic and Josephson inductance nonlinearities.*

### 2.3 Quantum noise and the standard quantum limit

Quantum amplifiers are fundamentally constrained by the uncertainty principle, which sets a lower bound on the simultaneous precision with which two conjugate quadratures of an electromagnetic mode can be defined. For a single mode with canonical quadratures $\hat{X}$ and $\hat{P}$, the uncertainty relation is:

$$\Delta X\,\Delta P \geq \frac{\hbar}{2}. \tag{6}$$

Here, $\Delta X$ and $\Delta P$ denote the standard deviations of the quadrature operators. This relation implies that any phase-preserving linear amplifier—which amplifies both quadratures equally—must inevitably add noise in order to satisfy the uncertainty bound [14].

The *Standard Quantum Limit* (SQL) defines this minimum added noise, corresponding to the equivalent of half a photon at the signal frequency. This sets the ultimate performance bound for any phase-preserving amplifier permitted by quantum mechanics. In terms of noise temperature, the SQL is:

$$T_{\mathrm{SQL}} = \frac{\hbar\omega}{2k_{\mathrm{B}}}, \tag{7}$$

where $\omega$ is the angular frequency of the signal and $k_{\mathrm{B}}$ is the Boltzmann's constant. For qubit readout frequencies in the 4–8 GHz range, this corresponds to a noise temperature on the order of 100 mK—roughly an order of magnitude lower than that of the best cryogenic HEMTs (1–5 K) [12, 34].

### 2.4 Phase-preserving vs. phase-sensitive amplification

Quantum amplifiers are generally classified into two categories: phase-preserving and phase-sensitive [14].

- *Phase-preserving amplifiers* amplify both quadratures of the input signal equally. To satisfy the uncertainty principle, they must add noise, with the SQL setting a minimum of half a photon of added noise power.

- *Phase-sensitive amplifiers* amplify one quadrature while attenuating the other. In this case, amplification can occur without adding noise to the amplified quadrature, enabling below-SQL performance in that quadrature at the expense of increased uncertainty in the conjugate one. This mechanism produces quadrature noise squeezing, which is valuable in quantum metrology and quantum error correction [35].

### 2.5 Gain and added noise

One key performance metric for an amplifier is its *power gain* $G$, defined as

$$G = \frac{P_{\text{out}}}{P_{\text{in}}}, \tag{8}$$

where $P_{\text{out}}$ and $P_{\text{in}}$ denote the output and input signal powers, respectively.

High gain alone is insufficient for quantum applications; the added noise must also be minimized. For phase-preserving amplifiers, the fundamental link between gain and noise is captured by the Haus-Caves limit [14], which states:

$$N_{\text{add}} \geq \frac{1}{2}\hbar\omega(G-1), \tag{9}$$

where $N_{\text{add}}$ denotes the added noise referred to the input, expressed as an equivalent energy at the signal frequency [14]. This inequality shows that although higher gain improves the signal-to-noise ratio (SNR) in subsequent stages, it cannot eliminate the intrinsic quantum noise penalty of phase-preserving amplification.

## 3. Superconducting parametric amplifiers

To approach the SQL, amplifiers must employ a non-dissipative nonlinear element to enable frequency mixing without introducing significant noise.

Parametric amplification is the process by which the amplitude of a weak signal is increased by transferring energy from a strong external pump through a nonlinear medium. Unlike conventional amplifiers that rely on active semiconductor elements such as transistors, parametric amplifiers use the time-dependent variation of a system parameter—such as inductance or capacitance—to mediate energy exchange [14].

The primary figure of merit in parametric amplification is the *gain*, which quantifies the increase in signal power. Parametric amplifiers can typically achieve high gain (20–30 dB) with minimal added noise. However, this performance comes at the expense of *bandwidth*, which is approximately inversely proportional to the gain due to the fundamental gain-bandwidth tradeoff [36].

For resonator-based parametric amplifiers, the gain-bandwidth tradeoff can be expressed as:

$$\sqrt{G}\,\Delta f \approx \kappa, \tag{10}$$

where $G$ is the power gain, $\Delta f$ is the amplifier bandwidth, and $\kappa$ is the resonator linewidth (its energy decay rate, in Hz). Thus, as the gain increases, the bandwidth narrows so that the product remains approximately constant.

### 3.1 Frequency mixing

When a strong pump tone at frequency $\omega_p$ and a weak signal at frequency $\omega_s$ are applied to a nonlinear medium—such as a superconducting resonator with kinetic or Josephson inductance—new frequency components, known as *idler* tones ($\omega_i$), are generated according to

$$\omega_i = \omega_p \pm \omega_s. \tag{11}$$

This process, called frequency mixing, originates from the time-dependent modulation of the system's reactive impedance and is a direct consequence of nonlinearity [37, 38].

In nonlinear mixing, the idler frequency enables efficient energy transfer from the pump to the signal. The pump provides energy that is shared between the signal and idler, ensuring energy conservation and maintaining phase coherence. This allows signal amplification without introducing extra noise from dissipative processes. Without the idler, the pump could not effectively transfer energy, and amplification would be largely ineffective [35, 38].

To better understand how nonlinear systems generate and manipulate new frequency components, it is useful to distinguish between two fundamental mechanisms—three-wave mixing and four-wave mixing—illustrated in **Figure 2**. These processes arise from different orders of nonlinearity and obey distinct energy-conservation relations [39, 40].

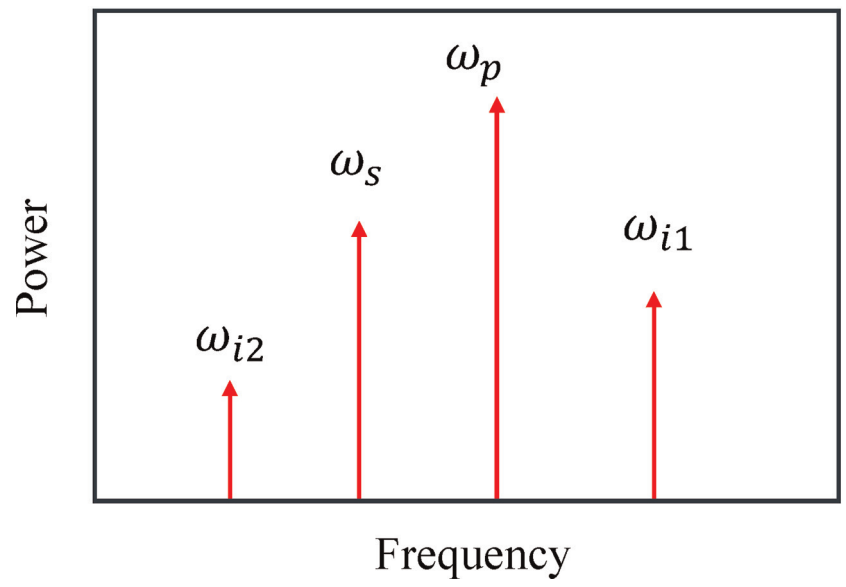

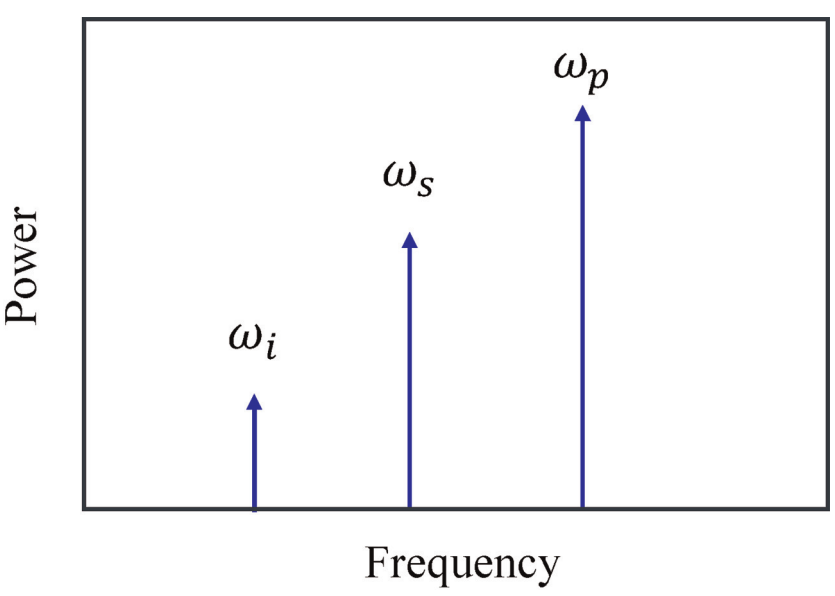

**Figure 2.**
*Three-wave and four-wave mixing. The pump ($\omega_p$) and signal ($\omega_s$) tones interact within a nonlinear resonant medium. (Left) In four-wave mixing, third-order nonlinearity generates two idler frequencies, $\omega_{i1} = 2\omega_p - \omega_s$ and $\omega_{i2} = 2\omega_s - \omega_p$, leading to frequency conversion and amplification. (Right) In three-wave mixing, second-order nonlinearity produces a single idler at $\omega_i = \omega_p - \omega_s$, also enabling frequency conversion and amplification. The idler power depends on both gain and pump strength.*

### 3.2 Degenerate vs. non-degenerate parametric amplification

As discussed above, parametric amplification is a nonlinear process in which energy from a pump tone is transferred to a signal and an idler. Depending on the frequency relationship between these tones, two main regimes arise: *degenerate*, where the signal and idler frequencies coincide, and *nondegenerate*, where they remain distinct [41, 42].

*Degenerate parametric amplification* occurs when the signal and idler frequencies coincide, that is, $\omega_s = \omega_i = \frac{\omega_p}{2}$. This requires both tones to lie within the same resonator mode, with the pump frequency approximately twice the resonator's natural frequency. Because the signal and idler share the same mode, the amplifier operates in a *phase-sensitive* regime, where the gain depends on the relative phase between the input signal and the pump. In this case, one quadrature of the signal can be amplified, while the orthogonal quadrature (90° out of phase) is attenuated. This enables noise squeezing below the Standard Quantum Limit (SQL), which is particularly valuable for quantum-limited measurements [43].

In contrast, *nondegenerate parametric amplification* occurs when the signal and idler occupy different modes of the circuit. These frequencies often couple through separate physical ports, each associated with its own resonance. In this configuration, the amplifier can provide (i) *reflection gain*, amplifying signals reflected at the same frequency, and (ii) *conversion gain*, transferring energy between the signal and idler frequencies [36, 43]. Unlike the degenerate case, nondegenerate amplification is largely phase-insensitive: It amplifies both quadratures equally and cannot achieve quadrature-selective gain or squeezing. Consequently, such amplifiers add at least half a photon of noise per quadrature [43].

### 3.3 Types of parametric amplifiers

In superconducting parametric amplifiers, signal amplification is achieved through wave-mixing processes driven by the device's intrinsic nonlinearity. This nonlinearity may arise from Josephson inductance (reactive) or from kinetic inductance, which can also involve dissipative contributions [44]. Accordingly, superconducting parametric amplifiers are broadly categorized into two main classes based on the underlying nonlinearity: Josephson Parametric Amplifiers (JPAs) [17, 45] and Kinetic Inductance Parametric Amplifiers (KIPAs) [38]. Both types can operate in either four-wave mixing (4WM) or three-wave mixing (3WM) regimes, depending on the form of nonlinearity used in the device.

#### *3.3.1 Josephson Parametric Amplifiers (JPAs)*

Josephson Parametric Amplifiers (JPAs) employ the intrinsic nonlinear inductance of Josephson junctions to achieve low-noise amplification of microwave signals. The effective inductance of a Josephson junction depends on the current flowing through it, allowing it to act as a nonlinear circuit element. The core principle is parametric amplification: A strong pump tone drives the junction into a nonlinear regime, periodically modulating the resonator's effective inductance. This modulation transfers energy from the pump to the signal, enabling amplification with minimal added noise (**Figure 3**) [45, 46].

JPAs operate through frequency mixing driven by a strong pump tone, with distinct regimes depending on circuit design and nonlinearity.

In the *three-wave mixing (degenerate) mode*, a pump at $\omega_p \approx 2\omega_s$ drives the circuit, producing signal ($\omega_s$) and idler ($\omega_i$) tones that overlap at the same frequency ($\omega_s = \omega_i$). This regime enables *phase-sensitive amplification*, where only one quadrature of the microwave field is amplified, ideally without introducing added noise. It is typically realized in *flux-pumped JPAs*, in which a Superconducting Quantum Interference Device (SQUID) terminates a resonator and its inductance is modulated by a flux drive at $2\omega_s$ [37, 47].

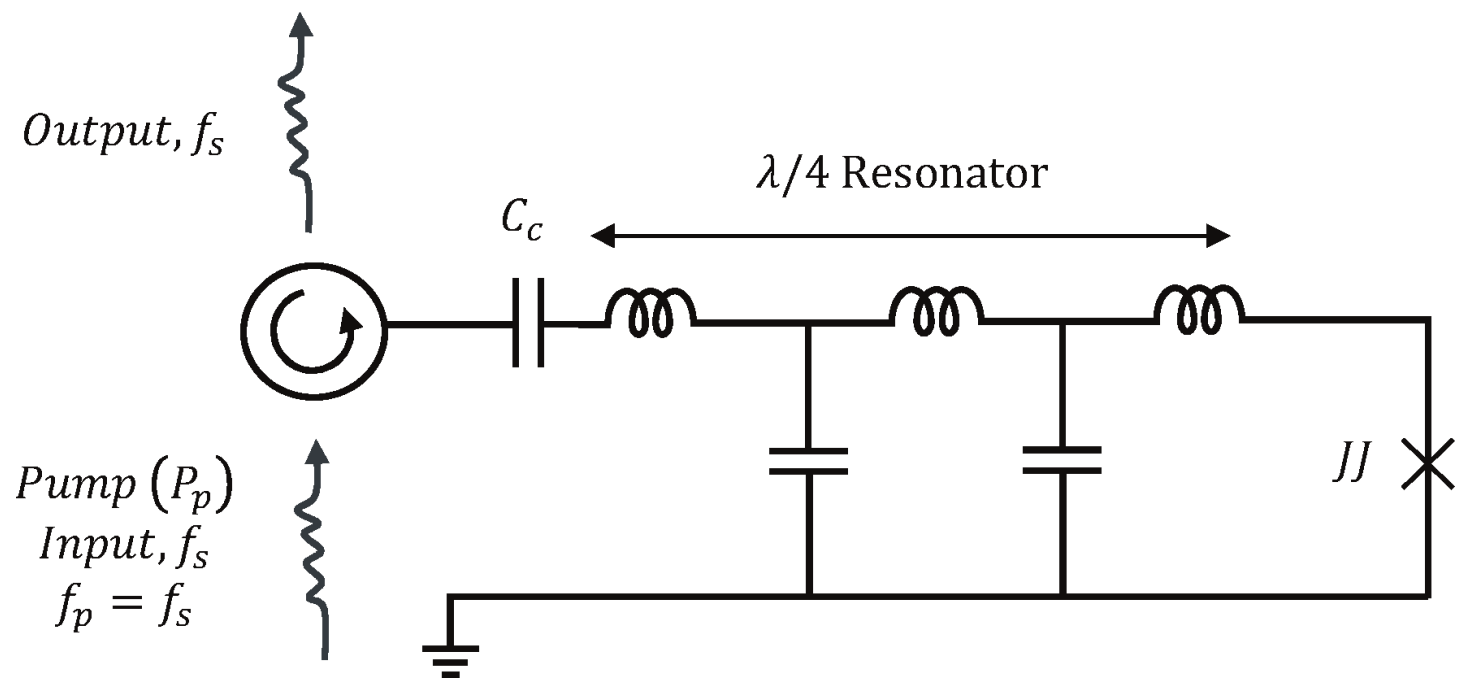

**Figure 3.**
*Lumped-element schematic representation of a resonant Josephson Parametric Amplifier (JPA). The device operates in reflection mode, with both the input signal ($f_s$) and pump tone ($P_p$) applied to the same port. A coupling capacitor ($C_c$) connects the input/output line to a $\lambda/4$ resonator modeled as a lumped LC network. At the grounded end, a Josephson junction (JJ) provides nonlinear inductance, enabling parametric amplification. In the degenerate three-wave configuration, the pump frequency is approximately twice the signal frequency ($f_p \approx 2f_s$). In practice, a circulator or isolator is placed at the port to separate the reflected amplified signal from the incoming tone.*

By contrast, in the *four-wave mixing (non-degenerate) mode*, the pump frequency is close to the signal frequency ($\omega_p \approx \omega_s$). Energy conservation requires $2\omega_p = \omega_s + \omega_i$, so the signal and idler emerge at distinct frequencies ($\omega_s \neq \omega_i$). This regime provides *phase-preserving amplification*, in which both quadratures are amplified equally. Such operation is commonly implemented in *current-pumped JPAs*, where a pump tone directly drives a single Josephson junction through a coupling capacitor [36, 43]. However, phase-preserving amplification is unavoidably constrained by the *Standard Quantum Limit*, introducing at least half a photon of added noise [14].

### *3.3.2 Traveling waveguide parametric amplifiers*

While JPAs provide near-quantum-limited amplification within a narrow bandwidth, their resonant nature inherently limits their frequency range and dynamic range [48]. To overcome these constraints, *Traveling Wave Parametric Amplifiers* (TWPAs) have been developed as a broadband alternative [49].

The operation of TWPAs is fundamentally based on the principle of parametric excitation, where energy is transferred to an oscillatory system by periodically varying one of its parameters, such as capacitance or inductance. For example, in a simple resonant circuit, if the capacitance is modulated in time—such as pulling apart condenser plates when the voltage reaches a maximum and restoring them when the voltage is zero—energy is added to the circuit, sustaining oscillations. This parametric modulation is typically driven at twice the resonant frequency of the circuit and can be implemented as a square wave or sinusoidal variation. In TWPAs, this concept is extended to a nonlinear transmission line, where the nonlinearity of Josephson junctions effectively modulates the system parameters dynamically as the pump tone propagates, enabling continuous energy transfer to the signal (**Figure 4**).

Unlike JPAs, which rely on a resonant structure, TWPAs use a nonlinear transmission line composed of a series of Josephson junctions or nonlinear elements that allow a pump tone to propagate along the line and continuously interact with the signal based on the principle of four-wave mixing to satisfy the frequency relation $2\omega_p = \omega_s + \omega_i$. This distributed nonlinearity facilitates parametric amplification over a wide frequency range without the bandwidth restrictions imposed by resonators [10, 38].

The continuous nature of the nonlinear medium allows the amplification to build up gradually over the entire length of the device, enabling high gain (often exceeding 20 dB) and large bandwidths (several GHz) simultaneously [50].

TWPAs typically employ arrays of Josephson junctions embedded in coplanar waveguides or lumped-element transmission lines, with careful impedance engineering to maintain low insertion loss and high dynamic range. Their broadband, low-noise performance has made them indispensable for high-fidelity readout of superconducting qubits and other quantum devices [10, 36].

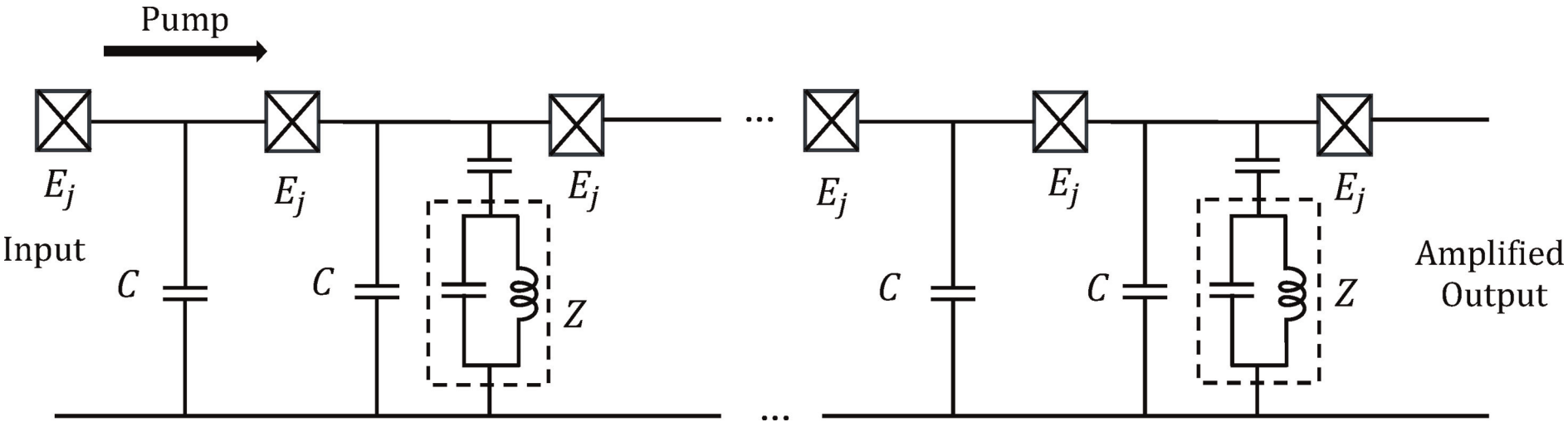

**Figure 4.**
*Schematic representation of a Traveling-Wave Parametric Amplifier (TWPA) based on a Josephson junction transmission line. The amplifier consists of a series of Josephson junctions (cross symbols) and capacitive shunt elements, with periodically inserted phase-matching structures (inductor-capacitor networks). The input signal enters on the left, is amplified through parametric interaction with a strong pump tone, and exits as an amplified output on the right.*

### *3.3.3 Kinetic Inductance Parametric Amplifiers*

Kinetic Inductance Parametric Amplifiers (KIPAs) are the key components in quantum information processing, especially for high-sensitivity quantum measurements. They are typically deployed as the first amplification stage, operating at the lowest temperature to provide sufficient gain and overcome noise from subsequent stages, such as high-electron-mobility-transistor (HEMT) amplifiers [51].

KIPAs operate based on the nonlinear kinetic inductance of thin superconducting films (e.g., NbTiN or NbN), which arises from the inertia of Cooper pairs. This nonlinearity becomes more pronounced in thin films and under current bias. Under an applied current, the kinetic inductance $L_k$ becomes current-dependent and is often approximated as:

$$L_k(I) \approx L_{k0}\left(1 + \alpha I^2\right), \tag{12}$$

where $L_{k0}$ is the zero-bias kinetic inductance and $\alpha$ characterizes the nonlinearity.

In these devices, four-wave mixing (4WM) can occur in transmission lines patterned as coplanar waveguides or microstrips when driven by a strong pump tone. Phase matching is required to maximize gain across the desired bandwidth, which is typically achieved through dispersion engineering. By tailoring the geometric dispersion, phase matching is maintained while suppressing unwanted higher-order processes, such as the pump's third harmonic.

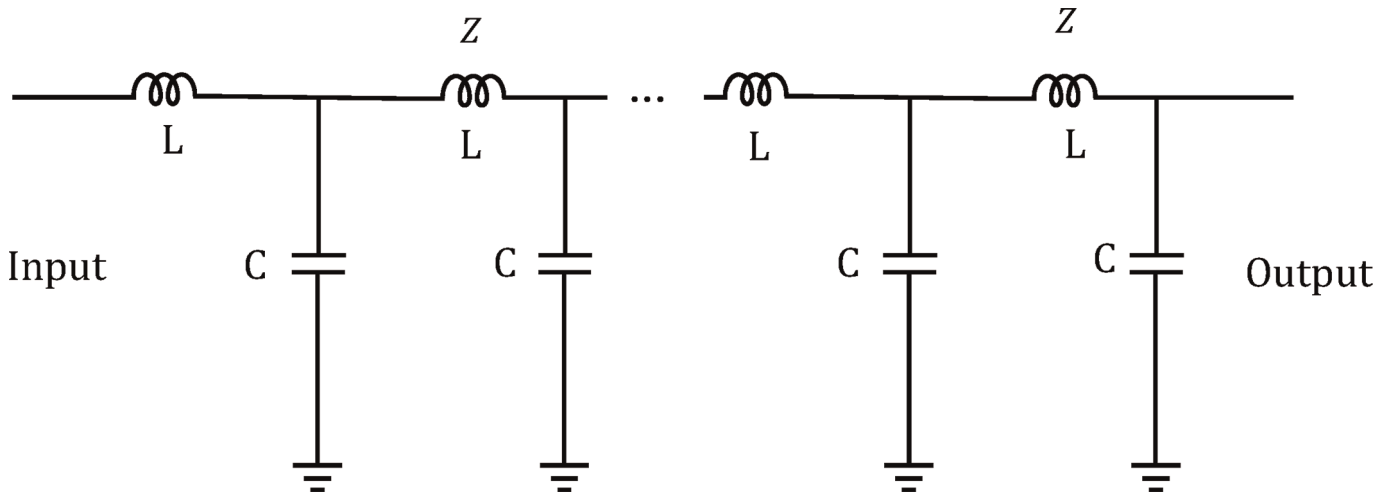

**Figure 5.**
*Layout of a Kinetic Inductance Parametric Amplifier (KIPA) based on a nonlinear transmission line. The structure consists of alternating coplanar waveguide (CPW) sections with engineered impedance and embedded kinetic inductance elements. The Z segments are cascaded to form a stepped-impedance profile, enabling phase matching and parametric gain. These segments can be designed as $\lambda/4$ or $\lambda/2$ CPWs.*

Applying a DC bias allows three-wave mixing (3WM) in kinetic inductance traveling-wave parametric amplifiers (KI-TWPAs), yielding near-quantum-limited noise performance, broad bandwidth, and high dynamic range (**Figure 5**) [52, 53].

Having explored the physical principles and architectural variations of parametric amplifiers—including JPAs, TWPAs, and KIPAs—it is essential to understand how these devices are integrated into a cryogenic measurement chain for qubit readout. The following schematic illustrates a typical experimental setup used in superconducting quantum circuits, where signal integrity, noise suppression, and thermal isolation are critical. Each amplifier type discussed earlier plays a distinct role depending on its placement, gain characteristics, and compatibility with the surrounding microwave infrastructure (**Figure 6**).

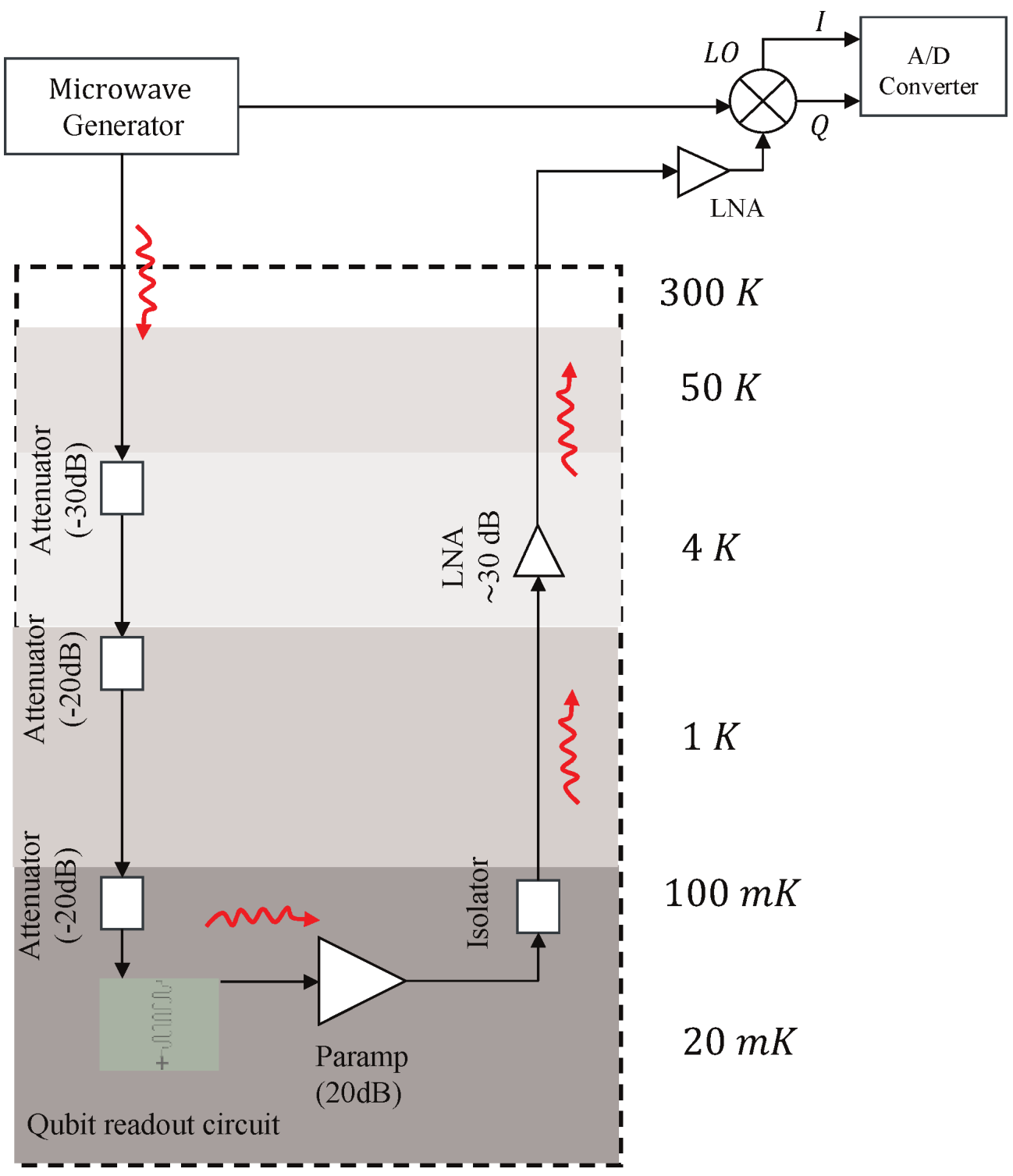

**Figure 6.**
*Cryogenic measurement setup for superconducting qubit readout: A probe signal from a vector network analyzer (VNA) passes through a series of attenuators at progressively colder temperature stages to suppress thermal noise. At the 20 mK stage, the signal interacts with the qubit circuit and is amplified by a parametric amplifier (Paramp) with $\sim$ 20 dB gain. An isolator at 100 mK prevents back-action noise from higher stages. The signal is further amplified by a low-noise amplifier (LNA) at 4 K and then demodulated* via *an I/Q mixer driven by a local oscillator (LO). The final signal is digitized by an analog-to-digital converter (A/D), enabling high-fidelity qubit state discrimination.*

## 4. Superconducting resonators: Structure and function in parametric amplifiers

### 4.1 Basics of microwave resonators

Microwave resonators are fundamental building blocks in superconducting parametric amplifiers (SPAs), providing energy storage and defining the interaction between microwave signals and nonlinear elements. A resonator confines electromagnetic energy and oscillates at discrete frequencies where constructive interference occurs—reflected waves combine in phase to reinforce each other [54]. In SPAs, the resonator enhances the pump tone and facilitates energy transfer between the pump, signal, and idler modes [38].

Superconducting resonators operate by forming standing waves at specific resonant frequencies. The use of superconducting materials is essential, as their unique properties, particularly the superconducting gap, ensure extremely low energy dissipation, yielding very high-quality factors. To minimize external disturbances such as thermal noise and stray radiation, these devices are typically operated at millikelvin temperatures inside dilution refrigerators [55, 56].

The main types of superconducting resonators are:

#### *4.1.1 Lumped-element (LC) resonators*

Lumped-element resonators consist of discrete inductors ($L$) and capacitors ($C$) connected in series or parallel to form an LC circuit. Their resonance frequency is given by:

$$f_0 = \frac{1}{2\pi\sqrt{LC}}. \tag{13}$$

These resonators provide a compact solution for low-frequency applications, where the physical size is much smaller than the operating wavelength [57]. At microwave frequencies, however, parasitic elements such as lead inductance, interturn capacitance, and conductor losses (including the skin effect) become significant and limit the achievable quality factor ($Q$). To overcome these limitations, modern designs employ superconducting thin films and integrated parallel-plate capacitors, reducing parasitics and achieving $Q$ factors exceeding $10^5$ while occupying an area of approximately 1 $mm^2$ [58].

#### *4.1.2 Distributed-element (CPW) resonators*

Distributed-element resonators, such as coplanar waveguide (CPW) resonators, are implemented using sections of transmission line designed to support standing-wave boundary conditions, typically $\lambda/4$ or $\lambda/2$ [59]. Unlike lumped-element designs, the inductance and capacitance in CPW resonators are distributed along the length of the line, eliminating the need for discrete components. The characteristic impedance $Z_0$ (commonly around 50 Ω) and propagation properties are determined by the inductance and capacitance per unit length of the transmission line [60].

While a CPW resonator's behavior at resonance can often be approximated by a lumped LC circuit for simplified analysis, its inherent distributed nature fundamentally governs all its resonant modes and frequency-dependent characteristics, both on- and off-resonance. CPW resonators are widely used in superconducting quantum circuits due to their ease of design up to frequencies of approximately 10 GHz, immunity to stray parasitics, and excellent compatibility with planar fabrication processes. When implemented using superconducting films at cryogenic temperatures, these resonators can achieve loaded quality factors ranging from a few hundred to several hundred thousand [59].

In CPW resonators, the electric field is concentrated in the gaps between the center conductor and the ground planes, with the magnetic field primarily surrounding the center conductor. This geometry allows precise control of impedance and coupling by adjusting the gap and conductor width. Engineering the field profile is essential for optimizing nonlinear participation in SPAs (**Figure 7**) [61].

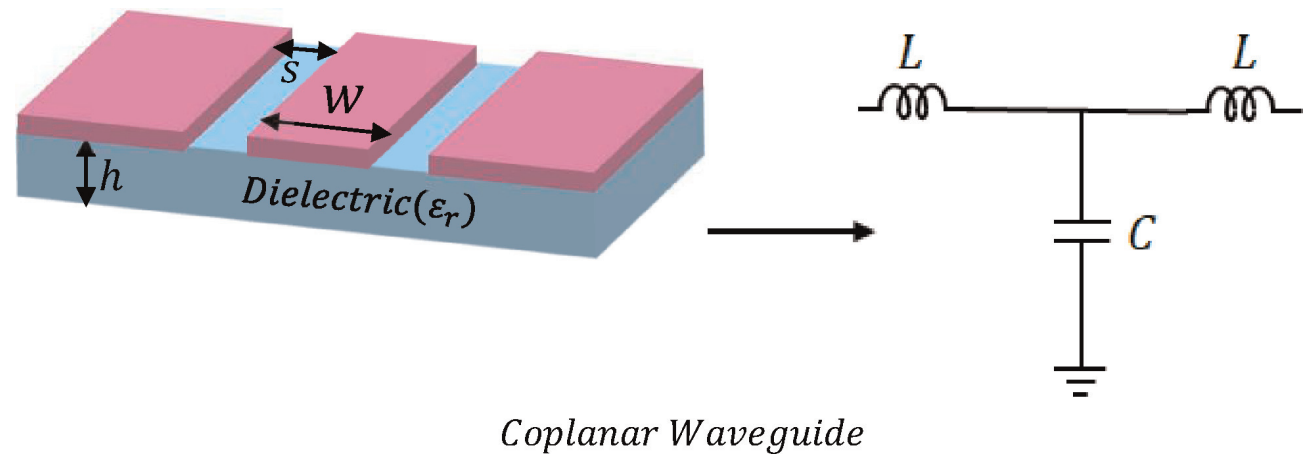

**Figure 7.**
*Schematic of a coplanar waveguide (CPW) resonator and its equivalent lumped-element model. The CPW consists of a central conductor of width W separated by gaps s from the ground planes on either side, all superconducting material patterned on a dielectric substrate of thickness h and relative permittivity $\varepsilon_r$.*

#### *4.1.3 $\lambda/4$ and $\lambda/2$ resonators*

Quarter-wave ($\lambda/4$) and half-wave ($\lambda/2$) configurations refer to the resonator length relative to the guided wavelength of the electromagnetic wave it supports. The guided wavelength is given by:

$$\lambda = \frac{v_p}{f} = \frac{c}{f\sqrt{\varepsilon_{\text{eff}}}}, \tag{14}$$

where $v_p$ is the phase velocity of the wave in the transmission line, and $\varepsilon_{\text{eff}}$ is the effective dielectric constant of the coplanar waveguide (CPW) structure [54].

A $\lambda/2$ resonator, often realized as a half-wavelength section of transmission line, supports a standing wave with voltage nodes (and current antinodes) at both ends for the fundamental mode. Its fundamental resonance occurs when the physical length satisfies:

$$l \approx \frac{\lambda}{2}, \quad f_0 = \frac{v_p}{2l}. \tag{15}$$

In contrast, a $\lambda/4$ resonator is a quarter-wavelength section, typically with one end short-circuited to ground and the other end open or capacitively coupled to the feed line. In this configuration, the shorted end forms a voltage node, while the open end forms a voltage antinode for the fundamental mode, and the resonance condition is:

$$l \approx \frac{\lambda}{4}, \quad f_0 = \frac{v_p}{4l}. \tag{16}$$

Quarter-wave resonators are preferred when easy coupling to feed lines is desired, while half-wave resonators are chosen for symmetric boundary conditions or multi-mode operation [54, 59].

### 4.2 Quality factor $(Q)$

The *quality factor* $(Q)$ quantifies how efficiently a resonator stores electromagnetic energy relative to its losses and characterizes the sharpness of its resonance. It also influences gain, bandwidth, and noise performance [54]. It is defined as

$$Q = 2\pi \frac{\text{Energy Stored}}{\text{Energy Lost per Cycle}} = \frac{\omega_0 W}{P_{\text{loss}}}, \tag{17}$$

where $\omega_0 = 2\pi f_0$ is the resonant angular frequency, $W$ the stored energy, and $P_{\text{loss}}$ the dissipated power per cycle. In terms of bandwidth $\Delta f$:

$$Q = \frac{f_0}{\Delta f}. \tag{18}$$

In superconducting CPW resonators, different loss mechanisms are quantified by distinct quality factors. The internal quality factor, $Q_{\text{int}}$, accounts for intrinsic losses such as conductor residual resistance, dielectric absorption, and radiation, and is defined by $Q_{\text{int}} = \frac{\omega_0 W}{P_{\text{internal}}}$. The external quality factor, $Q_{\text{ext}}$, describes losses due to intentional coupling to external circuitry, such as feed lines, and is given by $Q_{\text{ext}} = \frac{\omega_0 W}{P_{\text{external}}}$.

When both internal and external losses are present, the total or loaded quality factor, $Q_L$, satisfies $\frac{1}{Q_L} = \frac{1}{Q_{\text{int}}} + \frac{1}{Q_{\text{ext}}}$.

The loaded $Q$ determines the operational bandwidth and energy decay rate of the resonator.

## 5. Design and simulation of a CPW resonator-feed line system for qubit readout

In this section, we present the design and simulation of a planar superconducting readout circuit based on coplanar waveguide (CPW) resonators coupled to a common feed line.

This architecture, widely used in circuit quantum electrodynamics (cQED), enables efficient dispersive measurement of superconducting qubits in the microwave regime. We focus on a single-qubit configuration in which the qubit is coupled to a $\lambda/4$ meandered CPW resonator, and the resonator is interfaced with a CPW feed line for multiplexed readout. We outline the functional role of each circuit element, discuss key design considerations, and highlight how geometry sets parameters such as resonant frequency, coupling strength, and quality factors (see **Figure 8**).

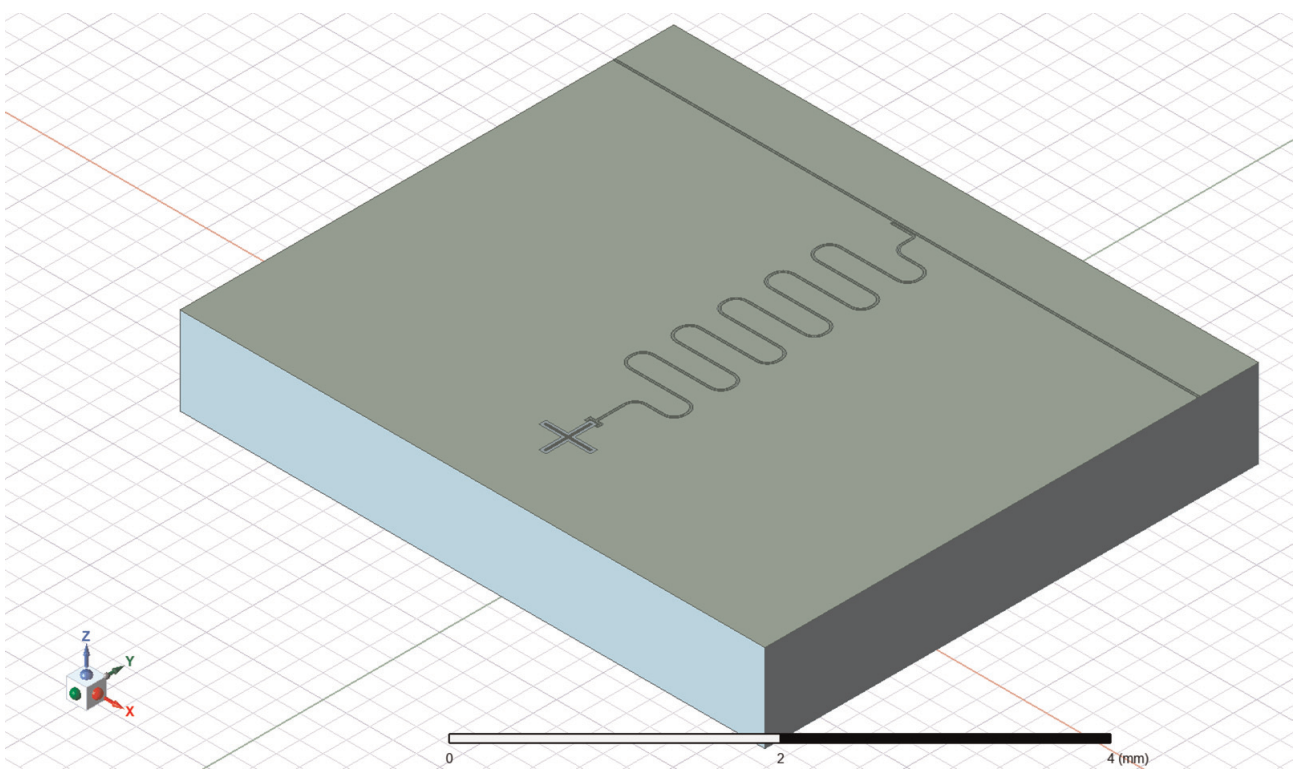

**Figure 8.**
*A qubit circuit using CPW as a resonator coupled with the feed line modeled in HFSS.*

## 5.1 Design overview

The primary application of this structure is the dispersive readout of superconducting qubits in the microwave domain, particularly for frequency multiplexing. In this scheme, multiple resonators of slightly different lengths are coupled to the same feed line, enabling each qubit to be read out at a distinct frequency.

The following is what each component's role is:

1. *Feed line:*

   The feed line is implemented as a coplanar waveguide (CPW), whose role is to deliver and collect microwave signals from the resonators. It transmits readout pulses to the resonators, collects the reflected or transmitted signals that encode qubit states, and also supports multiplexed readout when multiple resonators are coupled along its length.

   The first step in designing the coupling circuit is to realize a CPW feed line with a characteristic impedance of 50 Ω, ensuring proper matching to measurement equipment and minimizing reflections. The substrate is silicon with relative permittivity $\varepsilon_r = 11.7$ and thickness $h = 550, \mu$m. The CPW impedance is set by the conductor width $w$ and gap $s$, which are iteratively adjusted to achieve the desired impedance. The resulting optimized dimensions are illustrated in **Figure 7**.

   *Geometric definitions:*

   Here, $w$ is the width of the center conductor, $s$ the gap between the center conductor and the ground planes, $h$ the substrate thickness, and $\varepsilon_r$ the relative permittivity of the substrate.

   We define the normalized geometric ratios:

$$k_0 = \frac{w}{w + 2s}, \qquad k_1 = \frac{\sinh\left(\frac{\pi w}{4h}\right)}{\sinh\left(\frac{\pi(w+2s)}{4h}\right)} \tag{19}$$

*Elliptic integral approximations:*

The complete elliptic integrals of the first kind, $K(k_0)$ and $K(k_1)$, are approximated using logarithmic expressions depending on the value of $k$ [62]:

*For* $0 \le k \le 0.71$*:*

$$K(k) \approx \left[\frac{1}{\pi} \ln\left(\frac{2\left(1+\sqrt{\sqrt{1-k^2}}\right)}{1-\sqrt{\sqrt{1-k^2}}}\right)\right]^{-1} \tag{20}$$

*For* $0.71 < k \le 1$*:*

$$K(k) \approx \frac{1}{\pi} \ln\left(\frac{2\left(1+\sqrt{k}\right)}{1-\sqrt{k}}\right) \tag{21}$$

These approximations allow us to compute the ratio $K(k_1)/K(k_0)$, which governs both the impedance and the effective permittivity.

*Effective dielectric constant:*

The effective dielectric constant is given by:

$$\varepsilon_{\text{eff}} = 1 + \frac{\varepsilon_r - 1}{2} \cdot \frac{K(k_1)}{K(k_0)} \tag{22}$$

This accounts for the field distribution between the substrate and air regions.

*Characteristic impedance:*

The characteristic impedance of the CPW is:

$$Z_0 = \frac{30\pi}{\sqrt{\varepsilon_{\text{eff}}}} \cdot \frac{1}{K(k_0)} \tag{23}$$

This analytical expression assumes negligible conductor thickness and an ideal lossless dielectric. In practice, it does not capture the effects of finite metallization thickness, field interactions with air in open-back or suspended substrates, or deviations arising from fabrication tolerances and material imperfections. For accurate design, full-wave electromagnetic simulations or calibrated measurements are therefore recommended.

Considering the equations 19 to 21, for or case $w = 10\ \mu$m and $s = 6\ \mu$m, which yield approximately 50 Ω impedance (**Figure 9**).

2.*Meander:*

The meandered resonator functions as a frequency-selective element with a designed resonance in the 4–8 GHz range. It couples weakly to the feed line at one end and to the qubit at the other. In the dispersive regime, the qubit state induces a small shift in the resonator frequency, which can be monitored *via* the feed line to infer the qubit state non-destructively. A meandered geometry is employed to increase the effective resonator length while conserving chip area.

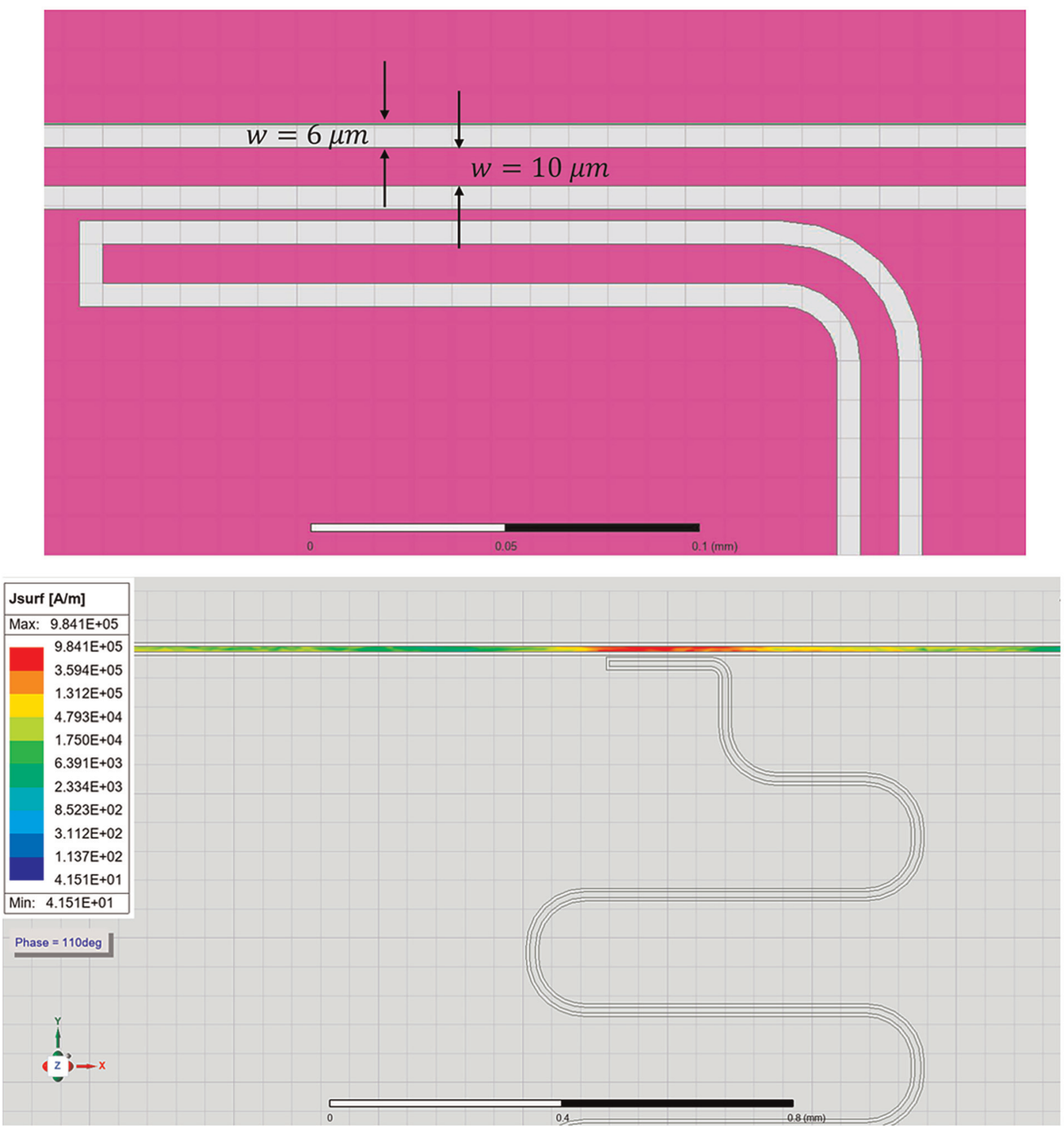

**Figure 9.**
*Feedline: (Top) feed line geometry (Bottom) simulated J-field related to feed line.*

The resonator is implemented as a $\lambda/4$ coplanar waveguide (CPW) resonator fabricated from a thin superconducting film, modeled as a perfect conductor in initial simulations. The conductor width $w$ and gap $s$ are chosen to target a characteristic impedance of 50 Ω, with dimensions ($w = 10\ \mu$m, $s = 6\ \mu$m) consistent with those of the feed line. To achieve the desired resonance frequency near 5.5 GHz while remaining compatible with typical chip dimensions, the resonator is realized as a quarter-wavelength transmission line. One end of the line is shorted to ground, while the opposite end is open and capacitively coupled to the transmon qubit. These boundary conditions yield a resonance frequency of

$$f_r \approx \frac{c}{4l\sqrt{\varepsilon_{\text{eff}}}}, \tag{24}$$

where $l$ is the resonator length (in our case $\sim 5.43$ mm), and $\varepsilon_{\text{eff}} = 6.3$ is the effective dielectric constant of the CPW, calculated from Eq. (23).

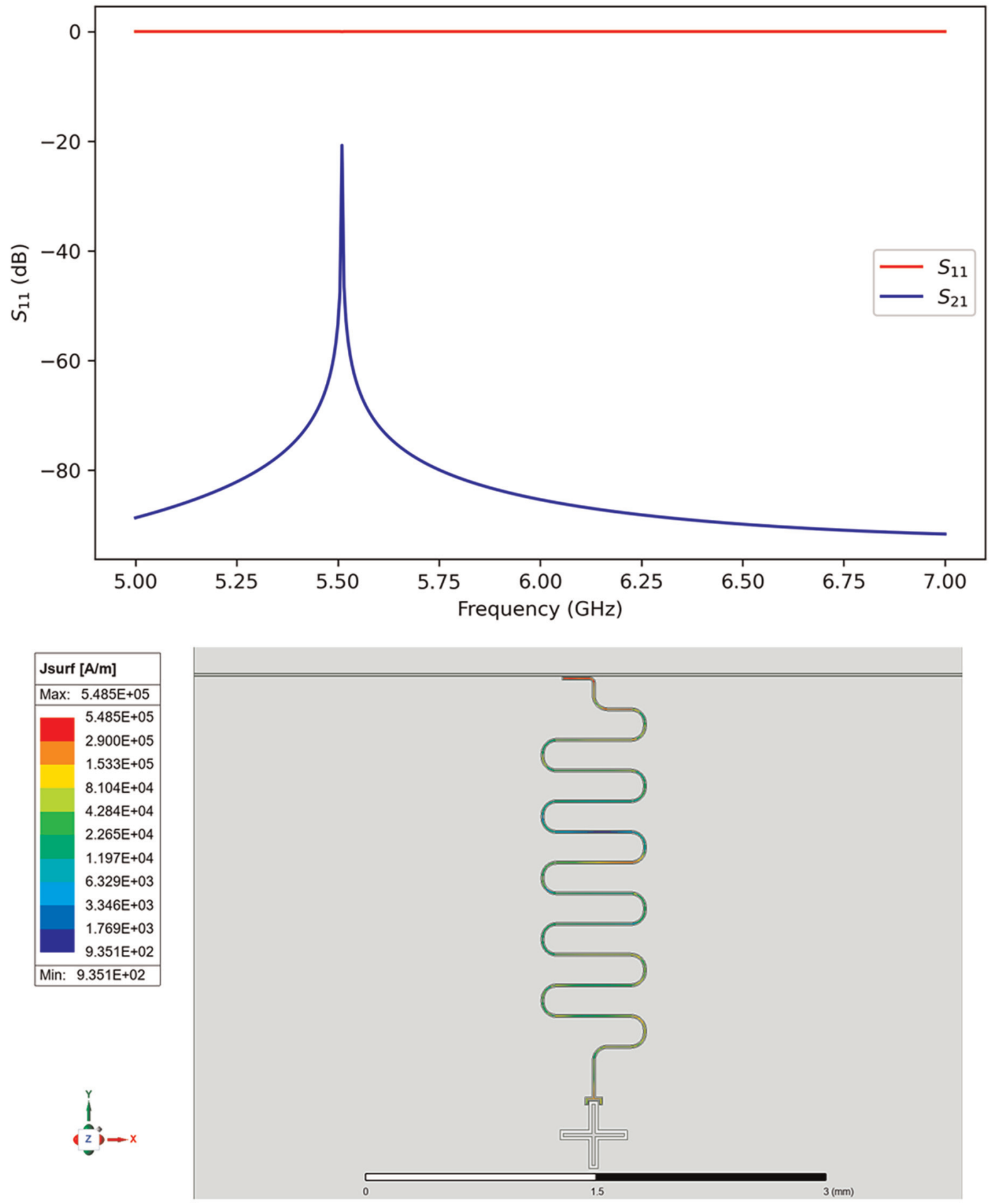

**Figure 10.**
*Meander resonating frequency and related field plot.*

3.*Cross-shaped structure:*

The cross-shaped capacitor pads of the transmon qubit provide large shunt capacitance to the Josephson junction, thereby reducing sensitivity to charge noise. They also couple to the resonator, enabling state readout and microwave control. The transmon qubit functions as a nonlinear oscillator, with a transition frequency determined by its Josephson energy ($E_J$).

*Coupling regions:*

- *Top short CPW gaps:*

  The feed line-resonator coupling region sets the external quality factor ($Q_c$) of the resonator. It governs the rate at which energy leaks from the resonator into the feed line, thereby influencing both readout speed and signal strength.

In practice, a CPW resonator is side-coupled to the feed line through a gap capacitor. This capacitance controls the external quality factor $Q_c$, and thus the coupling strength between the resonator and the measurement line. To calculate the effective capacitance, inductance, and mode frequencies, the CPW geometry (metal widths, spacings, and gaps) is transformed into a simpler equivalent geometry using conformal mapping techniques, where analytical integrals can be evaluated.

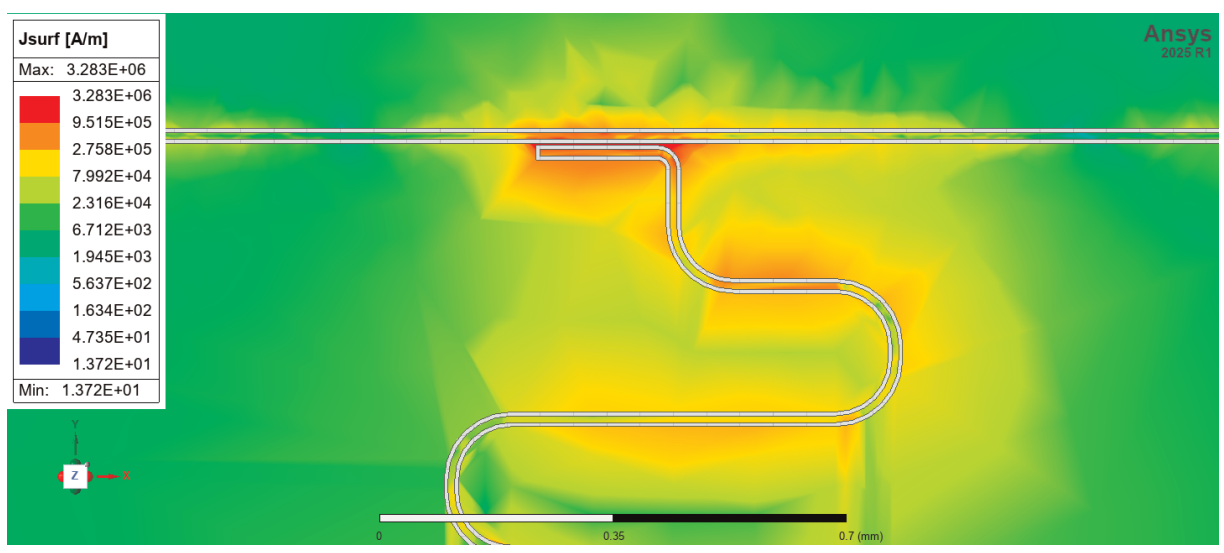

**Figure 11.**
*Field plot of the feed line coupling with meander.*

The resonator's electrical length sets the bare quarter-wave resonance frequency, $f_r^{(0)} = \frac{c_\ell}{4(l_o + l_c + l_s)}$, where $c_\ell = c/\sqrt{\varepsilon_{\text{eff}}}$ is the phase velocity on the line and $l_o$, $l_c$, and $l_s$ denote the open section, the coupling section, and the shorted section, respectively.

The presence of the feed line introduces an interaction described by a dimensionless coupling coefficient $\kappa$, which depends mainly on the coupling gap and overlap, and by two phase factors, $\theta = \frac{2\pi f_r^{(0)} l_c}{c_\ell}$, $\psi = \frac{2\pi f_r^{(0)} (l_c + 2l_o)}{c_\ell}$. These phases, together with $\kappa$, enter the first-order analytical expressions for the external quality factor $Q_e$ and the frequency shift of the coupled resonator [63].

In our design, considering the geometries shown in **Figures 10–12** and using the analytical description presented in [63], the feed line and resonator impedance are 49.81 and 49.94Ω, $\kappa = 0.093$ with the quality factor of $\sim 15000$.

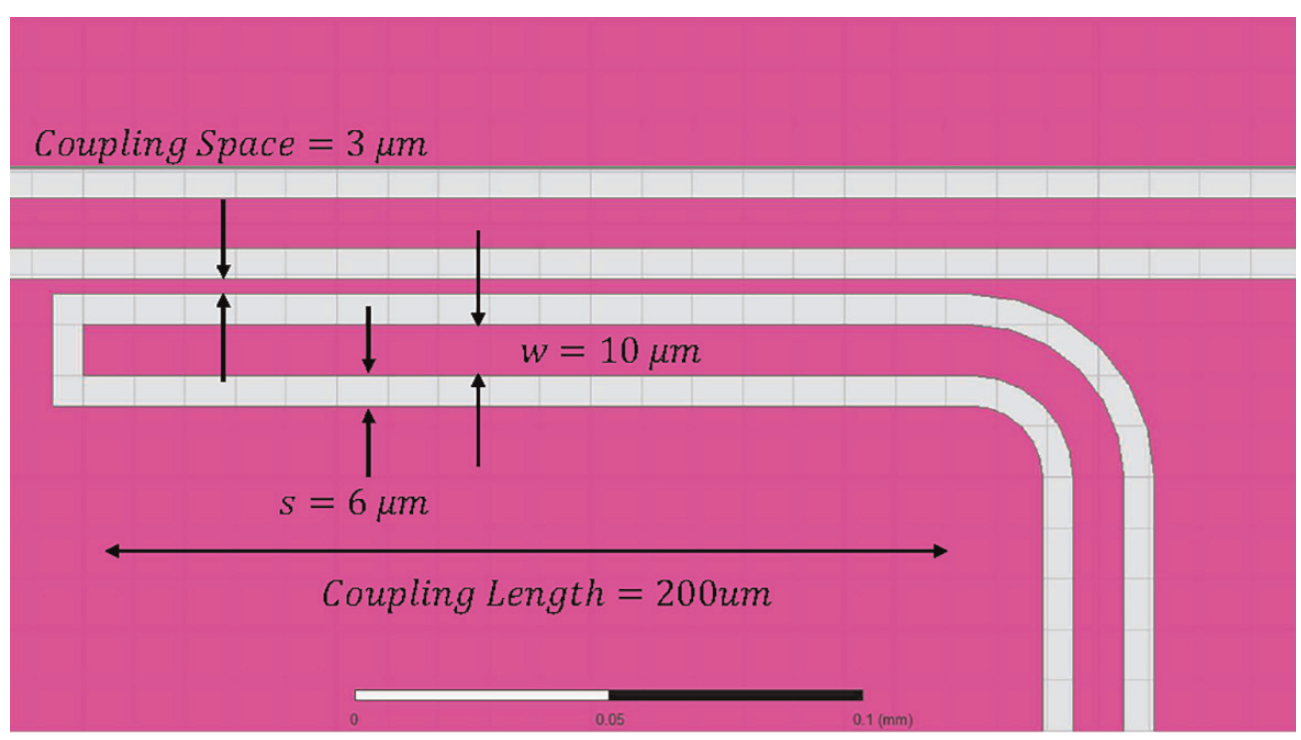

**Figure 12.**
*Meander and feed line coupling geometry.*

- *Bottom connection:*

  This region is the resonator-qubit coupling and sets the dispersive coupling rate $g$, influencing readout contrast and qubit-resonator interaction. To achieve this coupling the same calculation procedure is applied for this part as mentioned above (**Figure 13**).

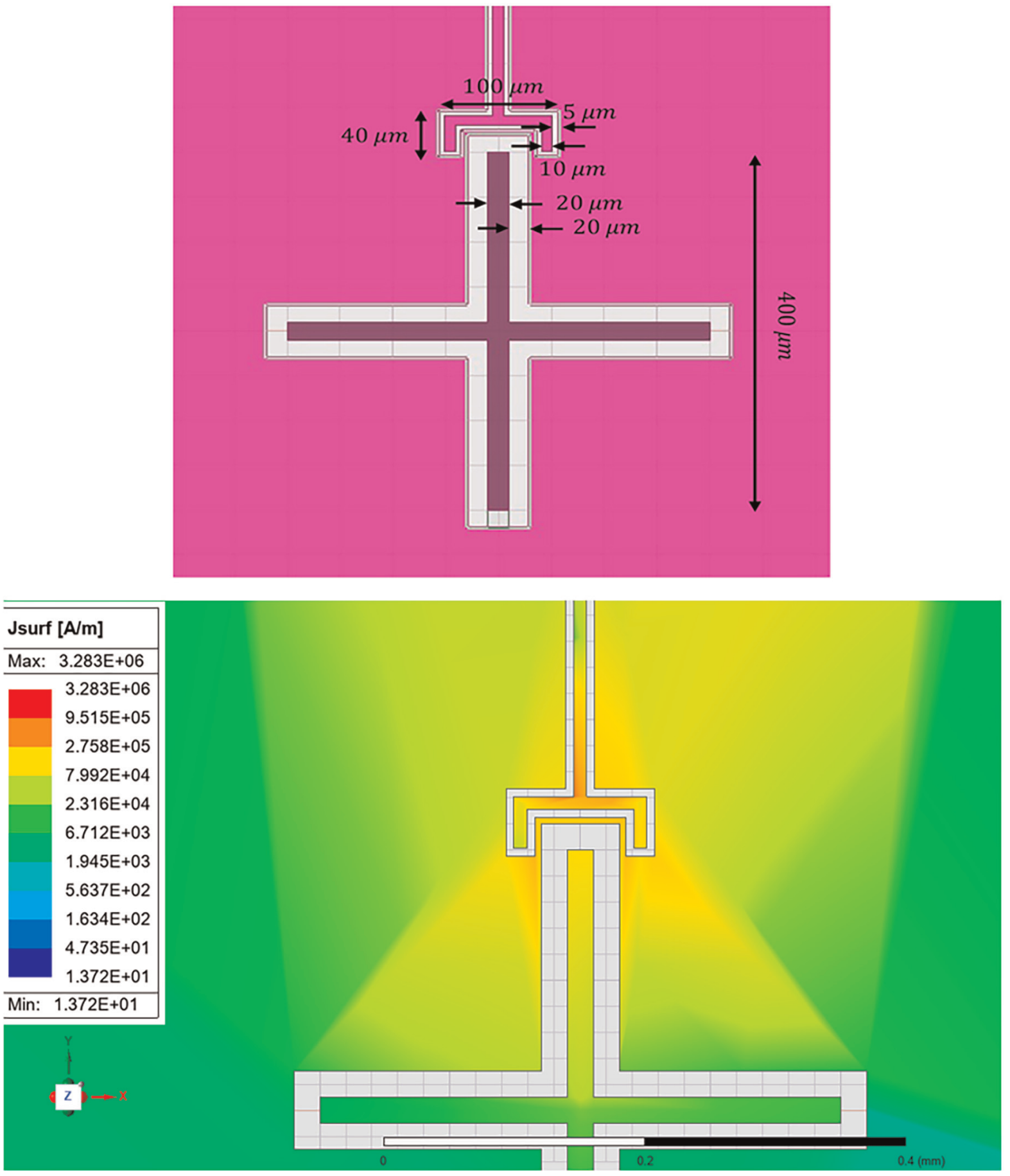

**Figure 13.**
*Cross-shaped capacitor coupling geometry.*

In this example, the coupling between the feed line and the resonator was implemented using a meander structure, and the resulting microwave response was analyzed to identify the resonance behavior of the readout, as shown in **Figure 14**.

This readout architecture, together with the analytical framework described above, can be naturally extended to multiple resonators. By introducing additional meandered branches along a single feed line, the system can enable simultaneous readout of several qubits in a scalable and modular fashion, while preserving independent control over coupling strengths and resonance frequencies.

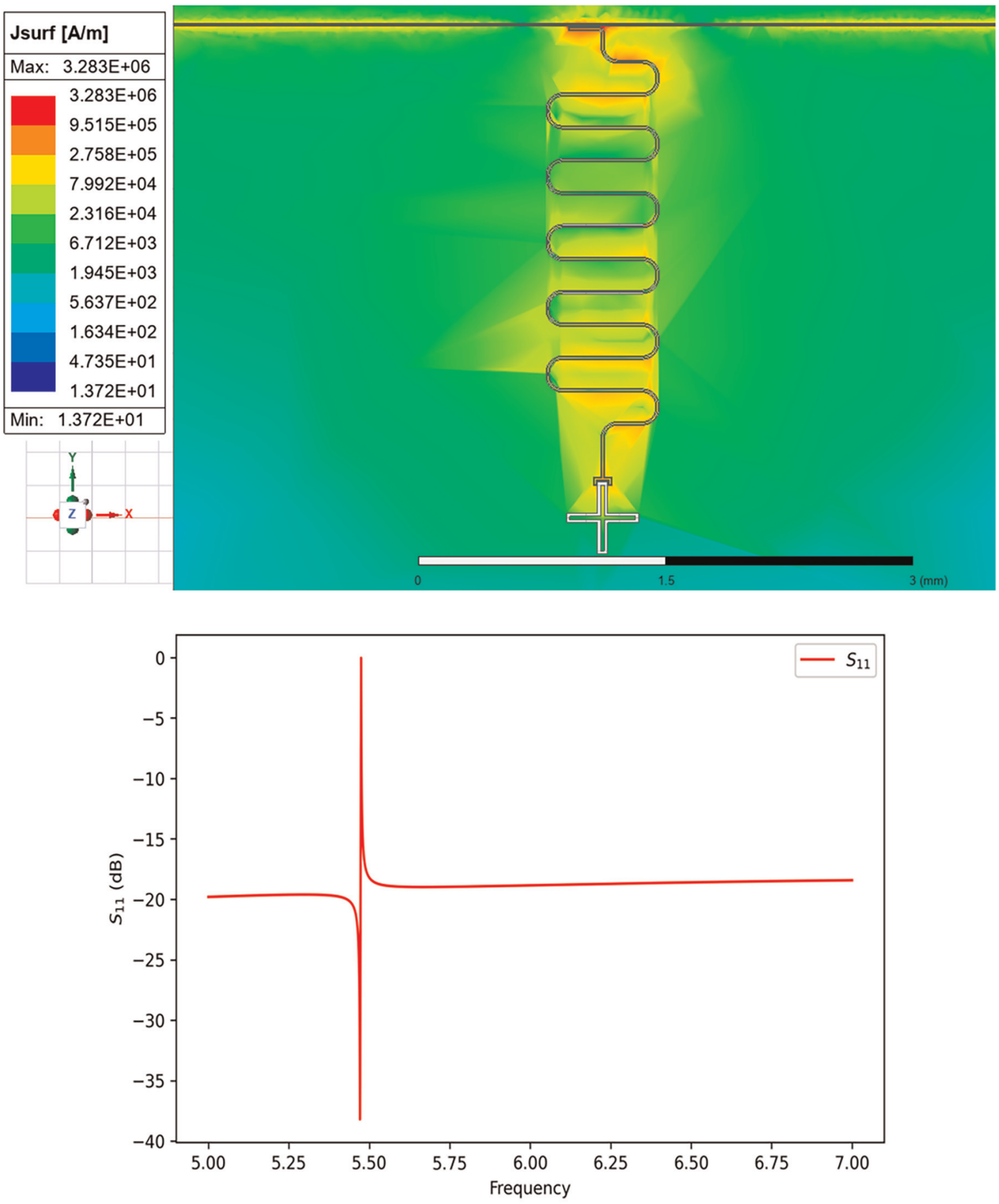

**Figure 14.**
*Design and response of a meander-coupled CPW resonator. (Top) Layout of the coplanar waveguide feed line with a meander section used to control the coupling rate. (Bottom) Simulated $S_{11}$ spectrum showing the resonance frequency, where the depth and linewidth of the dip provide information on the coupling strength and loaded quality factor of the resonator.*

## 6. Conclusion

In this chapter, we have comprehensively explored the foundational principles and practical applications of Superconducting Parametric Amplifiers (SPAs) within quantum computing. We highlighted the critical role of SPAs in achieving ultra-low-noise qubit readout, addressing the limitations of conventional amplifiers and approaching —or even surpassing—the Standard Quantum Limit (SQL).

The chapter reviewed the operating principles of parametric amplification, emphasizing the importance of superconducting nonlinearities from Josephson junctions and kinetic inductance, and distinguishing between phase-preserving and phase-sensitive amplification modes. Resonator design was shown to be central to SPA performance, with both lumped-element and coplanar waveguide structures enabling high-quality factors and optimal energy transfer.

We also examined key SPA architectures, including Josephson Parametric Amplifiers (JPAs), Traveling-Wave Parametric Amplifiers (TWPAs), and Kinetic Inductance Parametric Amplifiers (KIPAs), each offering specific advantages in gain, bandwidth, and dynamic range. Finally, a practical example of a meandered quarter-wavelength CPW resonator coupled to a feed line illustrated the principles of precise engineering for dispersive qubit readout and multiplexed measurement.

## Acknowledgements

This work was supported by the Science and Technology Facilities Council (STFC) under grant number ST/X001229/1.

## Author details

Babak Mohammadian
Jodrell Bank Centre for Astrophysics, University of Manchester, Manchester, UK

*Address all correspondence to: babak.mohammadian@manchester.ac.uk

## References

[1] Nachman B, Urbanek M, de Jong WA, Bauer CW. Unfolding quantum computer readout noise. npj Quantum Information. 2020;**6**:(1):84

[2] Salmanoglu A, Sirat VS. Design of ultra-low noise amplifier for quantum applications (qlna). Quantum Information Processing. 2024;**23**(3):91

[3] Tinkham M. Introduction to Superconductivity. 2nd ed. New York: McGraw-Hill; 1996

[4] Martinis JM. Superconducting phase qubits. Quantum Information Processing. 2009;**8**:81-103

[5] Houck AA, Türeci HE, Koch J. On-chip quantum simulation with superconducting circuits. Nature Physics. 2012;**8**:292-299

[6] Koch J, Yu TM, Gambetta J, Houck AA, Schuster DI, Majer J, et al. Charge-insensitive qubit design derived from the cooper pair box. Physical Review A. 2007;**76**(4):042319

[7] Krantz P, Kjaergaard M, Yan F, Orlando TP, Gustavsson S, Oliver WD. A quantum engineer's guide to superconducting qubits. Applied Physics Reviews. 2019;**6**(2):021318

[8] Clarke J, Wilhelm FK. Superconducting quantum bits. Nature. 2008;**453**(7198):1031-1042

[9] Renger J et al. Quantum-limited josephson traveling-wave parametric amplifier operating at the single-photon level. npj Quantum Information. 2021; **7**(1):89

[10] Macklin C, O'Brien K, Hover D, Schwartz ME, Bolkhovsky V, Zhang X, et al. A near-quantum-limited josephson traveling-wave parametric amplifier. Science. 2015;**350**(6258):307-310

[11] Mimura T. The early history of the high electron mobility transistor (hemt). IEEE Transactions on Microwave Theory and Techniques. 2002;**50**(3):780-782

[12] Choi AY, Esho I, Gabritchidze B, Kooi J, Minnich AJ. Characterization of self-heating in cryogenic high electron mobility transistors using schottky thermometry. Journal of Applied Physics. 2021;**130**(15):155107

[13] Mohammadian B, McCulloch MA, Sweetnam T, Gilles V, Piccirillo L. The impact of surface passivation on kapitza resistance at the interface between a semiconductor and liquid nitrogen. Journal of Low Temperature Physics. 2024;**214**(3):125-132

[14] Caves CM. Quantum limits on noise in linear amplifiers. Physical Review D. 1982;**26**(8):1817-1839

[15] Duh KHG, Pospieszalski MW, Kopp WF, Ho P, Jabra AA, Chao P-C, et al. Ultra-low-noise cryogenic high-electron-mobility transistors. IEEE Transactions on Electron Devices. 1988; **35**(3):249-256

[16] McKinstrie CJ, Radic S, Chraplyvy AR. Parametric amplifiers driven by two pump waves. IEEE Journal of Selected Topics in Quantum Electronics. 2002;**8**(3):538-547

[17] Aumentado J. Superconducting parametric amplifiers: The state of the art in josephson parametric amplifiers. IEEE Microwave Magazine. 2020;**21**(8): 45-59

[18] Wendin G. Quantum information processing with superconducting

circuits: A review. Reports on Progress in Physics. 2017;**80**(10):106001

[19] Patel PS, Desai DB. Review of qubit-based quantum sensing. Quantum Information Processing. 2025;**24**(3):1-37

[20] Kakande J. Phase sensitive parametric amplifiers and their applications [PhD thesis]. Southampton, United Kingdom: University of Southampton; 2012

[21] Lee W, Afshari E. Low-noise parametric resonant amplifier. IEEE Transactions on Circuits and Systems I: Regular Papers. 2010;**58**(3):479-492

[22] Lidar DA, Brun TA. Quantum Error Correction. Cambridge, United Kingdom: Cambridge University Press; 2013

[23] Degen CL, Reinhard F, Cappellaro P. Quantum sensing. Reviews of Modern Physics. 2017;**89**(3):035002

[24] Annunziata AJ, Santavicca DF, Frunzio L, Catelani G, Rooks MJ, Frydman A, et al. Tunable superconducting nanoinductors. Nanotechnology. 2010;**21**(44):445202

[25] Josephson BD. Possible new effects in superconductive tunnelling. Physics Letters. 1962;**1**:251-253

[26] Yariv A. Quantum Electronics. 3rd ed. New York, USA: John Wiley & Sons; 1989

[27] Roch N, Flurin E, Nguyen F, Morfin P, Campagne-Ibarcq P, Devoret MH, et al. Widely tunable, nondegenerate three-wave mixing microwave device operating near the quantum limit. Physical Review Letters. 2012;**108**(14):147701

[28] Meissner W, Ochsenfeld R. Ein neuer effekt bei eintritt der supraleitfähigkeit. Naturwissenschaften. 1933;**21**(44):787-788

[29] Bardeen J, Cooper LN, Schrieffer JR. Theory of superconductivity. Physical Review. 1957;**108**(5):1175

[30] London F, London H. The electromagnetic equations of the supraconductor. Proceedings of the Royal Society of London. Series A: Mathematical and Physical Sciences. 1935;**149**(866):71-88

[31] Tinkham M. Introduction to Superconductivity. North Chelmsford, MA, USA: Courier Corporation; 2004

[32] Vissers MR, Hubmayr J, Sandberg M, Chaudhuri S, Bockstiegel C, Gao J. Frequency-tunable superconducting resonators via nonlinear kinetic inductance. Applied Physics Letters. 2015;**107**(6):062601

[33] Kaiser W, Haider M, Russer JA, Russer P, Jirauschek C. Markovian dynamics of josephson parametric amplification. Advances in Radio Science. 2017;**15**:131-136

[34] Krantz P. Phonon black-body radiation limit for heat dissipation in electronics. ResearchGate. 2014. DOI: 10.13140/RG.2.1.1531.8487

[35] Clerk AA, Devoret MH, Girvin SM, Marquardt F, Schoelkopf RJ. Introduction to quantum noise, measurement, and amplification. Reviews of Modern Physics. 2010;**82**(2): 1155-1208

[36] Roy A, Devoret MH. Introduction to parametric amplification of quantum signals with josephson circuits. Comptes Rendus Physique. 2016;**17**(7):740-755

[37] Yamamoto T, Inomata K, Watanabe M, Matsuba K, Miyazaki T,

Oliver WD, et al. Flux-driven josephson parametric amplifier. Applied Physics Letters. 2008;**93**(4):042510

[38] Eom BH, Day PK, LeDuc HG, Zmuidzinas J. Wideband superconducting quantum interference device parametric amplifier. Nature Physics. 2012;**8**(8):623-627

[39] Agrawal GP. Nonlinear Fiber Optics. Boston, MA, USA: Academic Press; 2013

[40] Ernst M, Wulschner F, Lukashenko A, Marx A, Gross R. Kerr nonlinearity in superconducting circuits. Physical Review B. 2014;**90**(6):064508

[41] Parker DJ, Savytskyi M, Vine W, Laucht A, Duty T, Morello A, et al. Degenerate parametric amplification via three-wave mixing using kinetic inductance. Physical Review Applied. 2022;**17**(3):034064

[42] Winkel P, Takmakov I, Rieger D, Planat L, Hasch-Guichard W, Grünhaupt L, et al. Nondegenerate parametric amplifiers based on dispersion-engineered josephson-junction arrays. Physical Review Applied. 2020;**13**(2):024015

[43] Bergeal N et al. Phase-preserving amplification near the quantum limit with a josephson ring modulator. Nature. 2010;**465**(7294):64-68

[44] Zhao S, Withington S, Thomas CN. Superconducting resonator parametric amplifiers with intrinsic separation of pump and signal tones. Journal of Physics D: Applied Physics. 2024;**58**: 35305. Originally posted as arXiv: 2406.02455

[45] Yurke B, Kaminsky PG, Miller RE, Whittaker EA, Smith AD, Silver AH, et al. Observation of 4.2-k equilibrium-noise squeezing via a josephson-parametric amplifier. Physical Review Letters. 1988;**60**(9):764-767

[46] Devoret MH, Schoelkopf RJ. Superconducting circuits for quantum information: An outlook. Science. 2013; **339**(6124):1169-1174

[47] Castellanos-Beltran MA, Lehnert KW. Widely tunable parametric amplifier based on a superconducting quantum interference device array resonator. Applied Physics Letters. 2007; **91**(8):083509

[48] Castellanos-Beltran MA, Irwin KD, Vale LR, Hilton GC, Lehnert KW. Bandwidth and dynamic range of a widely tunable josephson parametric amplifier. IEEE Transactions on Applied Superconductivity. 2009;**19**(3): 944-947

[49] Grebel J, Clerk AA, Cleland AN. Design and performance of a flux-pumped impedance-engineered josephson parametric amplifier. Physical Review Applied. 2021;**16**(3):034046

[50] White TC et al. Traveling wave parametric amplifier with josephson junctions using minimal resonator loading. Applied Physics Letters. 2015; **106**(24):242601

[51] Splitthoff LJ, Wesdorp JJ, Pita-Vidal M, et al. Gate-tunable kinetic inductance parametric amplifier. Physical Review Applied. 2024;**21**(1):014029

[52] Malnou M, Mates JAB, Ranzani L, Vale LR, Hilton GC, Gao J, et al. Three-wave mixing kinetic inductance traveling-wave amplifier with near-quantum-limited noise performance. PRX Quantum. 2021;**2**:010302

[53] Klimovich NS. Traveling wave parametric amplifiers and other nonlinear kinetic inductance devices

[Ph.d. thesis]. Pasadena, CA, USA: California Institute of Technology; 2022

[54] Pozar DM. Microwave Engineering. Hoboken, NJ, USA: Wiley; 2011

[55] Leib M, Deppe F, Marx A, Gross R, Hartmann M. Networks of nonlinear superconducting transmission line resonators. New Journal of Physics. 2012; **14**:075024

[56] Pappas DP, Vissers MR, Wisbey DS, Kline JS, Gao J. Two level system loss in superconducting microwave resonators. In: IEEE Transactions on Applied Superconductivity, Conference Record of the 2010 Applied Superconductivity Conference. Vol. 21. Institute of Electrical and Electronics Engineers (IEEE); 2011. pp. 871-874

[57] Various. Superconducting microwave resonators: Design, fabrication, and applications. arXiv: 2109.07762. 2021

[58] Stockklauser AEA. Design of compact superconducting lc resonators with high quality factor. Physical Review Research. 2023;**5**(4):043126

[59] Göppl M, Fragner A, Baur M, Bianchetti R, Filipp S, Fink JM, et al. Coplanar waveguide resonators for circuit quantum electrodynamics. Journal of Applied Physics. 2008; **104**(11):113904

[60] National Institute of Standards and Technology (NIST). Design and characterization of coplanar waveguide resonators for quantum circuits. In: NIST Technical Note. U.S. Department of Commerce, National Institute of Standards and Technology (NIST); 2015. DOI: 10.6028/NIST.TN.1885

[61] Calusine G, Melville A, Braumüller J, Gudmundsen TJ, Oliver WD. Determining interface dielectric losses in superconducting resonators. Physical Review Applied. 2019;**12**:014012

[62] Wadell BC. Transmission Line Design Handbook. Norwood, MA: Artech House Publishers; 1991

[63] Besedin I, Menushenkov AP. Quality factor of a transmission line coupled coplanar waveguide resonator. EPJ Quantum Technology. 2018;**5**:2